\documentclass[12pt]{article}
\usepackage[english,german,french,polish]{babel}
\usepackage[T1]{fontenc}

\begin{document}
\selectlanguage{english}

\baselineskip 0.76cm
\topmargin -0.4in
\oddsidemargin -0.1in

\let\ni=\noindent

\renewcommand{\thefootnote}{\fnsymbol{footnote}}

\newcommand{\UK}{SuperKamiokande }

\newcommand{\CKM}{Cabibbo--Kobayashi--Maskawa }

\newcommand{\SM}{Standard Model }

\pagestyle {plain}

\setcounter{page}{1}

\pagestyle{empty}


~~~~~
\begin{flushright}
IFT-02/01 
\end{flushright}

\vspace{0.2cm}

{\large\centerline{\bf Search for fermion universality of the Dirac component}}
{\large\centerline{\bf of neutrino mass matrix{\footnote {Supported in part by the Polish State Committee for Scientific Research (KBN), grant 5 P03B 119 20 (2001--2002).}}}}

\vspace{0.5cm}

{\centerline {\sc Wojciech Kr\'{o}likowski}}

\vspace{0.23cm}

{\centerline {\it Institute of Theoretical Physics, Warsaw University}}

{\centerline {\it Ho\.{z}a 69,~~PL--00--681 Warszawa, ~Poland}}

\vspace{0.3cm}

{\centerline{\bf Abstract}}

\vspace{0.2cm}

An effective texture is presented for six Majorana neutrinos, three active and three (conventional) sterile, based on a 6x6 mass matrix, whose 3x3 active--sterile component ({\it i.e.}, Dirac component) is conjectured to get a {\it fermion universal form} similar to the constructed previously 3x3 mass matrix for charged leptons and 3x3 mass matrices for up and down quarks. This is true, however, {\it when} the bimaximal mixing, specific for neutrinos, is {\it transformed out unitarily} from the neutrino mass matrix. The 3x3 active--active component ({\it i.e.}, lefthanded component) of neutrino 6x6 mass matrix is diagonal and gets degenerate entries. It dominates over the whole neutrino mass matrix. In such a texture, three neutrino masses are nearly degenerate, $ m_1 \simeq m_2 \simeq m_3 $, but their mass-squared differences appear hierarchical, $\Delta m^2_{21} \ll \Delta m^2_{32} \simeq \Delta m^2_{31}$, while the remaining three neutrino masses can be constructed to vanish, $ m_4 = m_5 = m_6 = 0 $, in contrast to the familiar seesaw mechanism.

\vspace{0.3cm}

\ni PACS numbers: 12.15.Ff , 14.60.Pq , 12.15.Hh .

\vspace{0.6cm}

\ni January 2002

\vfill\eject

~~~~~
\pagestyle {plain}

\setcounter{page}{1}

\vspace{0.2cm}

\ni {\bf 1. Introduction}

\vspace{0.2cm}

As is well known, three Dirac neutrinos are $ \nu^{(D)}_\alpha = \nu_{\alpha L} + \nu_{\alpha R}\;\;(\alpha = e\,,\, \mu\,,\, \tau)$, while three Majorana active neutrinos and three Majorana (conventional) sterile neutrinos become $\nu^{(a)}_\alpha = \nu_{\alpha L} + \left( \nu_{\alpha L} \right)^c $ and $\nu^{(s)}_\alpha = \nu_{\alpha R} + \left(\nu_{\alpha R} \right)^c\;\;(\alpha = e\,,\, \mu\,,\, \tau)$, respectively. The neutrino mass term in the Lagrangian gets generically the form

\begin{equation} 
- {\cal L}_{\rm mass} = \frac{1}{2}\sum_{\alpha \beta} (\overline{\nu_\alpha^{(a)}} \,,\, \overline{\nu_{\alpha}^{(s)}}) \left( \begin{array}{cc} M^{(L)}_{\alpha \beta} & M^{(D)}_{\alpha \beta} \\ M^{(D)*}_{\beta \alpha}  & M^{(R)}_{\alpha \beta} \end{array} \right) \left( \begin{array}{c} \nu^{(a)}_\beta \\ \nu^{(s)}_{\beta} \end{array} \right) \;.
\end{equation} 
 
\ni If $M^{(L)}_{\alpha \beta} $ and $M^{(R)}_{\alpha \beta} $ are not all zero, then in nature there are realized six Majorana neutrino mass fields $\nu_i $ or states $|\nu_i \rangle \;\;(i = 1,2,3,4,5,6)$ connected with six Majorana neutrino flavor fields $\nu_\alpha $ or states $| \nu_\alpha\rangle \;\;(\alpha = e\,,\, \mu\,,\, \tau\,,\,e_s\,,\, \mu_s\,,\, \tau_s)$ through the unitary transformation
 
\begin{equation} 
\nu_\alpha  = \sum_i U_{\alpha i}  \nu_i \;\;{\rm or}\;\; |\nu_\alpha \rangle = \sum_i U^*_{\alpha i} |\nu_i \rangle \;,
\end{equation} 

\ni where we passed to the notation $\nu_\alpha \equiv \nu^{(a)}_\alpha $ and $\nu_{\alpha_s} \equiv \nu^{(s)}_\alpha $ for $\alpha = e\,,\, \mu\,,\, \tau$. Of course, $ \nu^{(a)}_{\alpha L} = \nu_{\alpha L} $, $\nu^{(a)}_{\alpha R} = (\nu_{\alpha L})^c $ and $\nu_{{\alpha_s} R} \equiv \nu^{(s)}_{\alpha R} = \nu_{\alpha R} $, $\nu_{{\alpha_s} L} \equiv \nu^{(s)}_{\alpha L} = (\nu_{\alpha R})^c $ for $\alpha = e\,,\, \mu\,,\, \tau$. Thus, the neutrino $6\times 6$ mass matrix
$M = (M_{\alpha \beta})\;\;(\alpha, \beta = e\,,\, \mu\,,\, \tau\,,\,e_s\,,\, \mu_s\,,\, \tau_s)$ is of the form 

\begin{equation}
M = \left( \begin{array}{cc} M^{(L)} & M^{(D)} \\ M^{(D)\dagger}  & M^{(R)} \end{array} \right) \;.
\end{equation}

The neutrino $6\times 6$ mixing matrix $U = (U_{\alpha i})\;\;(i = 1,2,3,4,5,6)$ appearing in Eqs. (2) is, at the same time, the unitary $6\times 6$ diagonalizing matrix,

\begin{equation} 
U^\dagger M U = M_{\rm d} \equiv {\rm diag}(m_1\,,\,m_2\,,\,m_3\,,\,m_4\,,\,m_5\,,\,m_6)\;,
\end{equation} 

\ni if the representation is used, where the charged-lepton $3\times 3$ mass matrix is diagonal. This will be assumed henceforth.

\vspace{0.2cm}

\ni {\bf 2. Model of neutrino texture}

\vspace{0.2cm}

In this paper we study the model of neutrino texture, where the $3\times 3$ submatrices in Eq. (3) are
 
\begin{equation} 
M^{(L)} = {\stackrel{0}{m}} \left( \begin{array}{ccc} 1 & 0 & 0 \\ 0 & 1 & 0 \\ 0 & 0 & 1 \end{array} \right) \,,\, M^{(D)} =  {\stackrel{0}{m}} \left( \begin{array}{ccc} \frac{t_{14}}{\sqrt{2}} & \frac{t_{25}}{\sqrt{2}} & 0 \\ -\frac{t_{14}}{2} &  \frac{t_{25}}{2} & \frac{t_{36}}{\sqrt{2}} \\ \frac{t_{14}}{2} & -\frac{t_{25}}{2} & \frac{t_{36}}{\sqrt{2}}  \end{array} \right) \,,\,
M^{(R)} = {\stackrel{0}{m}} \left( \begin{array}{ccc} t^2_{14} & 0 & 0 \\ 0 & t^2_{25} & 0 \\ 0 & 0 & t^2_{36} \end{array} \right)
\end{equation}

\ni with ${\stackrel{0}{m}}> 0 $ being a mass scale and $ t_{i j}\;\;(i j = 14\,,\,25\,,\,36)$ denoting three dimensionless parameters.

One can show that the unitary diagonalizing matrix $U$ for the mass matrix $M$ defined in Eqs. (3) and (5) is of the form

\begin{equation}
U = {\stackrel{1}{U}}{\stackrel{0}{U}}\;,\;{\stackrel{1}{U}} =  \left( \begin{array}{cc} U^{(3)} & 0^{(3)} \\ 0^{(3)} & 1^{(3)} \end{array} \right) \;,\; {\stackrel{0}{U}} = \left( \begin{array}{cc} C^{(3)} & -S^{(3)} \\ S^{(3)} & C^{(3)} \end{array} \right)\;,  
\end{equation} 

\ni where

\begin{eqnarray} 
U^{(3)} =   \left( \begin{array}{ccc} \frac{1}{\sqrt{2}} & \frac{1}{\sqrt{2}} & 0 \\ -\frac{1}{2} &  \frac{1}{2} & \frac{1}{\sqrt{2}} \\ \frac{1}{2} & -\frac{1}{2} & \frac{1}{\sqrt{2}}  \end{array} \right) & , &
1^{(3)} =  \left( \begin{array}{ccc} 1 & 0 & 0 \\ 0 & 1 & 0 \\ 0 & 0 & 1 \end{array} \right) ,\nonumber \\ C^{(3)} =  \left( \begin{array}{ccc} c_{14} & 0 & 0 \\ 0 & c_{25} & 0 \\ 0 & 0 & c_{36} \end{array} \right)\;\; & , &  S^{(3)} =  \left( \begin{array}{ccc} s_{14} & 0 & 0 \\ 0 & s_{25} & 0 \\ 0 & 0 & s_{36} \end{array} \right) 
\end{eqnarray} 

\ni and

\vspace{-0.2cm}

\begin{equation}
\frac{s_{ij}}{c_{ij}} = t_{ij} 
\end{equation}

\ni with $ s_{i j} = \sin \theta_{i j}$ and $ c_{i j} = \cos \theta_{i j}$, so that $ t_{i j} = \tan \theta_{i j}\;\;(ij = 14\,,\,25\,,\,36)$. Such a diagonalizing matrix leads to the mass spectrum

\begin{eqnarray}
m_1 = {\stackrel{0}{m}}\left( 1 + t^2_{14}\right) & , & m_4 = 0 \;, \nonumber \\
m_2 = {\stackrel{0}{m}}\left( 1 + t^2_{25}\right) & , & m_5 = 0 \;, \nonumber \\
m_3 = {\stackrel{0}{m}}\left( 1 + t^2_{36}\right) & , & m_6 = 0 
\end{eqnarray}

\ni which can be described equivalently by the equalities

\begin{equation}
c^2_{14} m_1 = c^2_{25} m_2 = c^2_{36} m_3 = {\stackrel{0}{m}}\;,\; m_4 = m_5 = m_6 = 0 \;.
\end{equation}

\ni The easiest way to prove this theorem is to start with the diagonalizing matrix $U$ given in Eqs. (6) and (7), and then to construct the mass matrix $M$ defined in Eqs. (3) and (5) by making use of the formula $M_{\alpha \beta} = \sum_i U_{\alpha i} m_i U^*_{\beta i}\,$, where the mass spectrum (9) or (10) is to be taken into account.

We can see from Eqs. (5), (6) and (7) that our neutrino texture corresponds to the mixing angles giving $ c_{12} = 1/\sqrt{2} = s_{12}$, $ c_{23} = 1/\sqrt{2} = s_{23} $ and $ c_{13} = 1$, $s_{13} = 0 $, while $ c_{ij} \,,\, s_{ij}\;\;(ij = 14\,,\,25\,,\,36)$ are to be determined from the experiment.

In this neutrino texture, where the mass matrix $M$ is given  in Eqs. (3) and (5), an interesting role is played by the unitarily transformed mass matrix ${\stackrel{0}{M}}$ defined as

\vspace{-0.2cm}

\begin{equation} 
{\stackrel{0}{M}} = {\stackrel{1}{U}\!^\dagger} M {\stackrel{1}{U}} \;.
\end{equation}

\ni Then, writing

\begin{equation}
{\stackrel{0}{M}} = \left( \begin{array}{cc} {\stackrel{0}{M}}^{(L)} & {\stackrel{0}{M}}^{(D)} \\ {\stackrel{0}{M}}^{(D)\dagger}  & {\stackrel{0}{M}}^{(R)} \end{array} \right) \;,
\end{equation}

\ni we obtain

\begin{eqnarray}
{\stackrel{0}{M}}\!\,^{(L)} = \left( {\stackrel{1}{U}\!^\dagger} M {\stackrel{1}{U}} \right)^{(L)}  & = & U^{(3) \dagger}M^{(L)} U^{(3)} = M^{(L)} \;,\nonumber \\
{\stackrel{0}{M}}\!\,^{(D)} = \left( {\stackrel{1}{U}\!^\dagger} M {\stackrel{1}{U}} \right)^{(D)}  & = & U^{(3) \dagger}M^{(D)} = {\stackrel{0}{m}}\left(\begin{array}{ccc} t_{14} & 0 & 0 \\ 0 & t_{25} & 0 \\ 0 & 0 & t_{36} \end{array}  \right) \;, \nonumber \\ {\stackrel{0}{M}}\!\,^{(R)} = \left( {\stackrel{1}{U}\!^\dagger} M {\stackrel{1}{U}} \right)^{(R)}  & = & M^{(R)}\;.
\end{eqnarray}

\ni Thus, the Dirac $3\times 3$ component ${\stackrel{0}{M}}\!\,^{(D)}$ of the mass matrix ${\stackrel{0}{M}}$ (transformed unitarily from $M$ by means of the factor ${\stackrel{1}{U}}$ of the mixing matrix $U$) becomes diagonal and so, may get a {\it hierarchical} structure {\it similar} to the Dirac mass matrices for charged leptons and quarks, all dominated by their diagonal parts. The transforming factor ${\stackrel{1}{U}}$ given in Eq. (6) works effectively thanks to its $3\times 3$ submatrix $U^{(3)}$ that is just the familiar {\it bimaximal mixing matrix} [1], specific for neutrinos, describing satisfactorily the observed oscillations of solar $\nu_e$'s and atmospheric $\nu_\mu $'s. Note that


\begin{equation} 
{\stackrel{0}{U}}^\dagger {\stackrel{0}{M}} {\stackrel{0}{U}} = {\stackrel{0}{M}}_d = M_d = {\rm diag}(m_1\,,\,m_2\,,\,m_3\,,\,m_4\,,\,m_5\,,\,m_6)\;,
\end{equation}

\ni where the factor ${\stackrel{0}{U}}$ of the mixing matrix $U$ is defined in Eq. (6). Here, ${\stackrel{0}{M}} = \left( {\stackrel{0}{M}}_{ij} \right)$ , ${\stackrel{0}{U}} = \left( {\stackrel{0}{U}}_{ij}\right)$ and ${\stackrel{1}{U}} = \left( {\stackrel{1}{U}}_{\alpha i}\right)$, as $M = (M_{\alpha \beta})$ and $U = (U_{\alpha i})$. With the use of ${\stackrel{0}{M}}$ given in Eq. (11) the neutrino mass term (1) in the Lagrangian can be written as $- {\cal L}_{\rm mass} = \frac{1}{2}\sum_{\alpha \beta} \overline{\nu}_\alpha M_{\alpha \beta} {\nu_\beta} = \frac{1}{2}\sum_{i j} \overline{\stackrel{0}{\nu}}_i {\stackrel{0}{M}}_{ij} {\stackrel{0}{\nu}}_j $, where $\nu_\alpha = \sum_i U_{\alpha i} \nu_i = \sum_i {\stackrel{1}{U}}_{\alpha i} \stackrel{0}{\nu}_i $, but $\stackrel{0}{\nu}_i = \sum_j {\stackrel{0}{U}}_{ij} \nu_j$ are not neutrino mass fields, in contrast to $\nu_i$: in fact, ${\stackrel{0}{M}} |{\stackrel{0}{\nu}}_i \rangle = m_i |{\stackrel{0}{\nu}}_i \rangle $, while $ M|{\nu}_i \rangle = m_i|{\nu}_i \rangle $ ( ${\stackrel{0}{M}} $  being a unitary transform of the full neutrino mass matrix $M$).

Specifically, the Dirac $3\times 3$ component ${\stackrel{0}{M}}\!\,^{(D)}$ of the neutrino mass matrix ${\stackrel{0}{M}}$ (where the bimaximal mixing characteristic for neutrinos is transformed out unitarily) may be conjectured in a {\it fermion universal form} that was shown to work very well for the mass matrix of charged leptons [2] and neatly for  mass matrices of up and down quarks [3] (obviously, in those three cases of charged fundamental fermions there exist only Dirac-type mass matrices). Then, for neutrinos we get [4]


\begin{equation}
{\stackrel{0}{M}}\!\,^{(D)} = \frac{1}{29} \left(\begin{array}{ccc} \mu \varepsilon & 2\alpha  & 0 \\ & & \\ 2\alpha  & 4\,\mu\, (80 + \varepsilon)/9 & 8\sqrt{3}\, \alpha \\ & & \\ 0 & 8\sqrt{3}\, \alpha & 24\, \mu\, (624 + \varepsilon)/25 \end{array} \right) \;,
\end{equation}

\ni where $\mu > 0$ , $\alpha > 0$ and $\varepsilon > 0$ are some neutrino parameters. Since already for charged leptons $\varepsilon^{(e)} = 0.172329$ is small [2], we will put for neutrinos $\varepsilon \rightarrow 0$. We will also conjecture that for neutrinos $\alpha/\mu $ is negligible, as for charged leptons the small $\left( \alpha^{(e)}/\mu^{(e)} \right)^2 = 0.023^{+0.029}_{-0.025} $ [2] gives the prediction $ m_\tau = m^{\rm exp}_\tau = 1777.03^{+0.30}_{-0.26}$ MeV [5] when $m_e = m^{\rm exp}_e$ and $m_\mu = m^{\rm exp}_\mu $ are used as inputs, while with $\left( \alpha^{(e)}/\mu^{(e)} \right)^2 = 0 $ the prediction becomes $ m_\tau = 1776.80$ MeV. In such a case, from Eqs. (13) and (15) we can conclude that


\begin{equation}
{\stackrel{0}{m}}\,t_{14} = \frac{\mu}{29} \varepsilon \rightarrow 0 \;,\;{\stackrel{0}{m}}\,t_{25} = \frac{\mu}{29} \frac{4\cdot 80}{9}  = 1.23 \,\mu \;,\; {\stackrel{0}{m}}\,t_{36} = \frac{\mu}{29} \frac{24 \cdot 624}{25} = 20.7 \,\mu
\end{equation}

\ni in Eqs.  (5), (8) and (9), and

\vspace{-0.3cm}

\begin{eqnarray}
c_{14} & \rightarrow & 1 \,,\, c_{25} = \frac{1}{\sqrt{1+1.50\, (\mu/\!\stackrel{0}{m})^2}} \,,\, c_{36} = \frac{1}{\sqrt{1+427\, (\mu/\!\stackrel{0}{m})^2}} \,,  \nonumber \\
s_{14} & \rightarrow & 0 \,,\, s_{25} = \frac{1.23\,\mu/{\stackrel{0}{m}}}{\sqrt{1+1.50\, (\mu/\!\stackrel{0}{m})^2}} \,,\, s_{36} = \frac{20.7\,\mu/{\stackrel{0}{m}}}{\sqrt{1+427\, (\mu/\!\stackrel{0}{m})^2}} 
\end{eqnarray}

\ni in Eqs. (7) and (8). Hence, from Eqs. (9) and (16) 

\begin{equation} 
m_1 \rightarrow {\stackrel{0}{m}}\,,\, m_2 = {\stackrel{0}{m}} + 1.50 \frac{\mu^2}{\stackrel{0}{m}} \,,\, m_3 = {\stackrel{0}{m}} + 427 \frac{\mu^2}{\stackrel{0}{m}}\,.
\end{equation} 

\vspace{0.2cm}

\ni {\bf 3. Neutrino oscillations}

\vspace{0.2cm}

Accepting the formulae (16) and making tentatively the conjecture that $\mu \ll {\stackrel{0}{m}}$, we can operate with the approximation, where $ 0 \leq t_{ij} \ll 1$ or $ 0 \leq s_{ij} \ll c_{ij} \;\; (ij = 14\,,\,25\,,\,36)$. Then, we get the case of nearly degenerate spectrum of $m_1\,,\,m_2\,,\,m_3 $: $m_1 \simeq m_2 \simeq m_3 \simeq {\stackrel{0}{m}}$, but with hierarchical mass-squared differences $ \Delta m_{21}^2 \ll \Delta m_{32}^2  \simeq \Delta m_{31}^2 $, where

\begin{eqnarray}
\Delta m^2_{21} & = & 2{\stackrel{0}{m}}\!\,^2\, (t^2_{25} - t^2_{14}) = 3.01 \, \mu^2 \;, \nonumber \\
\Delta m^2_{32} & = & 2{\stackrel{0}{m}}\!\,^2\, (t^2_{36} - t^2_{25}) = 850 \,\; \mu^2 \;, \nonumber \\
\Delta m^2_{31} & = & 2{\stackrel{0}{m}}\!\,^2\, (t^2_{36} - t^2_{14}) = 853 \,\; \mu^2 
\end{eqnarray}

\ni due to Eqs. (9) and (16).

Notice that the option ${\stackrel{0}{m}} \ll \mu $, opposite to our conjecture $\mu \ll {\stackrel{0}{m}}$, leads to $ t_{ij} \gg 1$ or $ 0 \leq c_{ij} \ll s_{ij}$ $(ij = 14,25,36)$. Then, we obtain the case of hierarchical spectrum of $ m_1\,,\,m_2\,,\,m_3 $: $m_1 \ll m_2 \ll m_3 $ with mass-squared differences $ \Delta m_{21}^2 \ll \Delta m_{32}^2  \simeq \Delta m_{31}^2 $, where 

\begin{eqnarray*}
\Delta m^2_{21} & = & {\stackrel{0}{m}}\!\,^2\, (t^4_{25} - t^4_{14}) = 2.26\, \frac{\mu^4}{ {\stackrel{0}{m}}\!\,^2} \,, \\
\Delta m^2_{32} & = & {\stackrel{0}{m}}\!\,^2\, (t^4_{36} - t^4_{25}) = 1.82 \times 10^5 \,\frac{\mu^4}{  {\stackrel{0}{m}}\!\,^2}\,, \\
\Delta m^2_{31} & = & {\stackrel{0}{m}}\!\,^2\, (t^4_{36} - t^4_{14}) = 1.82 \times 10^5 \,\frac{\mu^4}{ {\stackrel{0}{m}}\!\,^2}
\end{eqnarray*}

\vspace{-1.65cm}

\begin{flushright}
(19')
\end{flushright}

\vspace{0.3cm}

\ni due to Eqs. (9) and (16). In this case, the component $ M^{(R)}$ of the neutrino mass matrix dominates over $ \,M^{(D)}$ (as $\mu\, $ over $\,\stackrel{0}{m}\,$) that dominates in turn over $\, M^{(L)}$ (as  $\mu\,$ over $\,\stackrel{0}{m}\,$): this is the situation, where the familiar seesaw mechanism can formally work in spite of the fact that entries of $ M^{(R)}$ are very small, in particular due to $ m_4\, = m_5\, = m_6\, = 0 $ (not as in the popular seesaw, where they are as large as the GUT scale). With the Super-Kamiokande result $ \Delta m_{32}^2 \sim 3\times 10^{-3}\;\,{\rm eV}^2$ we get in this option $ \Delta m_{21}^2 \sim 3.7\times 10^{-8}\;\,{\rm eV}^2$ and ${\mu^4}/{\stackrel{0}{m}}\!\,^2 \sim 1.6\times 10^{-8}\;{\rm eV}^2$ or $\mu^2/{\stackrel{0}{m}} \sim 1.3\times 10^{-4}\;\,{\rm eV}$ {\it i.e.}, $\mu \sim 1.3\times 10^{-4}(\stackrel{0}{m}/\mu) \;\,{\rm eV} \ll 1.3\times 10^{-4}\;\,{\rm eV}$. In contrast, in the case of our conjecture $\mu \ll \stackrel{0}{m} $, the component $ M^{(L)}$ dominates over $ M^{(D)}$ which dominates in turn over $ M^{(R)}$ and so, we obtain for $ \Delta m_{21}^2$ and $\mu $ the much larger values given later on in Eqs. (26) and (24), respectively; also the value of $ \Delta m_{25}^2 = m^2_2 \simeq \,\stackrel{0}{m}\!^2 $ appearing in Eq. (28) is much larger.   

The familiar formulae for probabilities of neutrino oscillations $\nu_\alpha \rightarrow \nu_\beta $ on the energy shell,

\begin{equation} 
P(\nu_\alpha \rightarrow \nu_\beta) = |\langle \nu_\beta| e^{i PL} |\nu_\alpha \rangle |^2 = \delta _{\beta \alpha} - 4\sum_{j>i} U^*_{\beta j} U_{\beta i} U_{\alpha j} U^*_{\alpha i} \sin^2 x_{ji} 
\end{equation}

\ni with

\begin{equation} 
x_{ji} = 1.27 \frac{\Delta m^2_{ji} L}{E} \;,\; \Delta m^2_{ji}  = m^2_j - m^2_i \;,\, p_i \simeq E -\frac{m^2_i}{2E} \,,
\end{equation} 

\ni valid when a possible CP violation can be ignored (then $ U_{\alpha i}^* = U_{\alpha i}$), give in the accepted approximation of $ \Delta m_{21}^2 \ll \Delta m_{32}^2  \simeq \Delta m_{31}^2 \ll {\stackrel{0}{m}}\!\,^2 $ that

\begin{eqnarray} 
P(\nu_e \rightarrow \nu_e)_{\rm sol}\;\;\;\, & = & 1 -  c^2_{25} \sin^2 (x _{21})_{\rm sol} - \frac{1}{2} (1+c_{25}^2) s^2_{25} \;, \nonumber \\
P(\nu_\mu \rightarrow \nu_\mu)_{\rm atm}\;\; & = & 1 - \frac{1}{2}(1 + c^2_{25})c^2_{36} \sin^2 (x _{32})_{\rm atm} - \frac{1}{8}(1 + c^2_{25} + 2 c^2_{36})(s^2_{25} + 2 s^2_{36})\;, \nonumber \\
P(\nu_\mu \rightarrow \nu_e)_{\rm LSND} & = & \frac{1}{2} s^4_{25}\sin^2 (x _{25})_{\rm LSND}\;, \nonumber \\
P(\bar{\nu}_e \rightarrow \bar{\nu}_e)_{\rm Chooz} & = & 1 - (1 + c^2_{25})s^2_{25} \sin^2 (x _{25})_{\rm Chooz}
\end{eqnarray} 

\ni for solar $\nu_e$'s, atmospheric $\nu_\mu$'s, LSND accelerator $\nu_\mu$'s ($ \bar{\nu}_\mu $'s) and Chooz reactor $\bar{\nu}_e$'s, respectively. The first two Eqs. (22) differ from the familiar two--flavor oscillation formulae (used often in analyses of neutrino oscillations) by some additive terms that, fortunately, are small enough because of $ s^2_{ij} \ll c^2_{ij} $ consistent with $\mu^2 \ll \,\stackrel{0}{m}\!^2 $. 

 From the second formula (22) decribing atmospheric $\nu_\mu$'s we infer due to the SuperKamiokande result [6] that

\begin{equation} 
\frac{1}{2}(1 +c^2_{25}) c^2_{36} \equiv \sin^2 2\theta_{\rm atm} \sim 1 \;,\; \Delta m^2_{32} \equiv \Delta m^2_{\rm atm} \sim 3 \times 10^{-3}\;\,{\rm eV} \;,
\end{equation} 

\ni what gives 

\begin{equation} 
\mu^2 \sim 3.5 \times 10^{-6} \; {\rm eV}^2\;\, {\rm or}\,\;\mu \sim 1.9 \times 10^{-3} \; {\rm eV}  \;,
\end{equation} 

\ni when Eq. (19) is used. The nearly maximal atmospheric oscillation amplitude $ \sin^2 2 \theta_{\rm atm} \sim 1$ implies $c^2_{25} \sim 1$ and $c^2_{36} \sim 1$, which is consistent with $\mu^2 \ll \,\stackrel{0}{m}\!^2 $. For an illustration, taking $ \sin^2 2 \theta_{\rm atm} \stackrel{>}{\sim} 0.85$, we get from Eqs. (17) $(\mu/{\stackrel{0}{m}})^2 \stackrel{<}{\sim} 4.1 \times 10^{-4}$ and so, $\stackrel{0}{m}\!^2 \stackrel{>}{\sim} 8.3 \times 10^{-3}\;\;{\rm eV}^2 $ or ${\stackrel{0}{m}} \stackrel{>}{\sim} 9.3 \times 10^{-2}\;\;{\rm eV}$ due to Eq. (24). Thus, $ \sin^2 2 \theta_{\rm atm}$ should be much larger than 0.85 in order to have $\stackrel{0}{m}\!^2 \gg \Delta m^2_{32} \sim 3 \times 10^{-3}\;\;{\rm eV}^2 $. If {\it e.g.} $\stackrel{0}{m}\; \sim 1$ eV, then $ \sin^2 2 \theta_{\rm atm} \sim 0.998$.

Making use of the estimate (24) in Eqs. (18) we obtain

\begin{equation} 
m_1 \rightarrow {\stackrel{0}{m}}\,,\, m_2 \sim {\stackrel{0}{m}} + 5.3\times 10^{-6}\, \frac{{\rm eV}^2 }{\stackrel{0}{m}}\,,\, m_3 \sim {\stackrel{0}{m}} + 1.5\times 10^{-3}\, \frac{{\rm eV}^2}{\stackrel{0}{m}} \,.
\end{equation} 

The first formula (22) referring to solar $\nu_e$'s {\it predicts} with the use of Eqs. (19) and (24) that

\begin{equation} 
\sin^2 2\theta_{\rm sol} \equiv c^2_{25} \sim 1 \;\,,\,\;  
\Delta m^2_{\rm sol} \equiv \Delta m^2_{21} \sim 3.01 \mu^2 \sim 1.1 \times 10^{-5}\;\,{\rm eV}^2 \;.
\end{equation} 

\ni Such a prediction for solar $\nu_e$'s is not inconsistent with the Large Mixing Angle (LMA) solution [7], though the solar oscillation amplitude in this solution seems to be a bit smaller than the SuperKamiokande atmospheric oscillation amplitude (in contrast to the inequality $c^2_{25} > \frac{1}{2}(1 + c^2_{25})c^2_{36}$, where $c^2_{25}>c^2_{36}$ due to Eqs. (17); however the small additive terms $\frac{1}{2}(1 + c^2_{25})s^2_{25} < \frac{1}{8}(1 + c^2_{25} + 2c^2_{36})(s^2_{25} + 2s^2_{36})$ may compensate effectively such an inequality).

From the third formula (22) we can see that in our texture there is {\it predicted} a very small version of the original LSND effect for accelerator $\nu_\mu$'s ($ \bar{\nu}_\mu $'s)  [8] with the oscillation amplitude

\begin{equation} 
\sin^2 2 \theta_{\rm LSND} \equiv \frac{1}{2} s^4_{25} = \frac{1.13 (\mu/{\stackrel{0}{m}})^4}{[1 + 1.50 (\mu/{\stackrel{0}{m}})^2]^2} \sim 1.4 \times 10^{-11}\,\left(\frac{{\rm eV}}{\stackrel{0}{m}} \right)^4\,, 
\end{equation} 

\ni where Eqs. (17) and (24) are used (with $ \mu^2 \ll \,{\stackrel{0}{m}}\!\,^2$). The mass-squared scale for such a version of the LSND effect is equal to

\begin{equation} 
\Delta m^2_{\rm LSND} \equiv \Delta m^2_{25} = m^2_2  = {\stackrel{0}{m}}\!\,^2 + 3.01 \mu^2 \sim \,{\stackrel{0}{m}}\!\,^2 + 1.1\times 10^{-5}\, {\rm eV}^2 \;,
\end{equation} 

\ni where Eq. (19) is applied. Note that  $\Delta m^2_{\rm LSND}$ differs by the term $\,{\stackrel{0}{m}}\!\,^2\,$ from the solar mass-squared scale $\,\Delta m^2_{\rm sol}\,$ given in Eq. (26). If {\it e.g.} $\,{\stackrel{0}{m}} = O(10^{-1}\;{\rm eV})\, - \,O(1\;{\rm eV})$ (still with $ \,\mu^2 \ll \,{\stackrel{0}{m}}\!\,^2 $), then $\sin^2 2 \theta_{\rm LSND} = O(10^{-7}) - O(10^{-11})$ and $\Delta~ m^2_{\rm LSND} = O(10^{-2}\;{\rm eV}^2) - O(1\;{\rm eV}^2)$. 

The fourth formula (22) describes the Chooz experiment for reactor $\bar{\nu}_e$'s. Due to its negative result, $P(\bar{\nu}_e \rightarrow \bar{\nu}_e)_{\rm Chooz} \sim 1$, there appears the experimental constraint for $s^2_{25}$ [9]:

\begin{equation} 
(1 +c^2_{25})s^2_{25} \equiv \sin^2 2\theta_{\rm Chooz} \stackrel{<}{\sim} 0.1\;\;{\rm if}\;\;
\Delta m^2_{25} \equiv \Delta m^2_{\rm Chooz} \stackrel{>}{\sim} 0.1\,{\rm eV}^2\,.
\end{equation} 

\ni This implies for the LSND effect (in our texture) the Chooz upper bound  

\begin{equation} 
\sin^2 2 \theta_{\rm LSND} \equiv \frac{1}{2} s^4_{25}  \stackrel{<}{\sim} 1.3 \times 10^{-3}
\end{equation} 

\ni if $\Delta m^2_{25} \gg \Delta m^2_{32} \sim 3 \times 10^{-3}\;\,{\rm eV}^2 $, what is consistent with $\Delta m^2_{25} \stackrel{>}{\sim}0.1\;\,{\rm eV}^2 $ and gives $ (x_{25})_{\rm Chooz} \gg(x_{32})_{\rm Chooz} \simeq (x_{32})_{\rm atm} = O(1)$ as $(x_{ji})_{\rm Chooz} \simeq (x_{ji})_{\rm atm}$ numerically. Then, 

\begin{equation} 
\sin^2(x_{25})_{\rm Chooz} \simeq \frac{1}{2} 
\end{equation} 

\ni in the fourth formula (22). When combined with Eq. (27), the Chooz bound (30) leads to the lower limit for $\stackrel{0}{m}$:

\vspace{-0.1cm}

\begin{equation} 
\stackrel{0}{m}\, \stackrel{>}{\sim} 1.0 \times 10^{-2} \;{\rm eV}\;.
\end{equation} 

\ni This gives in turn the lower limits 

\vspace{-0.2cm}
\begin{equation} 
\Delta m^2_{\rm LSND} \equiv \Delta m^2_{25} \stackrel{>}{\sim}  1.1 \times 10^{-4}\;{\rm eV}^2
\end{equation} 

\ni and

\vspace{-0.1cm}

\begin{equation} 
\sin^2 2 \theta_{\rm sol} \equiv c^2_{25} \stackrel{>}{\sim}  0.95 \;\;,\;\;\sin^2 2 \theta_{\rm atm}  \equiv \frac{1}{2}(1 + c^2_{25}) c^2_{36} \stackrel{>}{\sim}  0.061
\end{equation} 

\vspace{0.15cm}

\ni due to Eqs. (17) and (28), respectively. Evidently, this lower limit for $\sin^2 2 \theta_{\rm atm}$ is not reached experimentally. If {\it e.g.} ${\stackrel{0}{m}} \sim 1$ eV corresponding to $\sin^2 2 \theta_{\rm atm} \sim 0.998 $, then $\sin^2 2 \theta_{\rm LSND} \sim 1.4\times \!10^{-11}\!$, $\Delta m^2_{\rm LSND} \sim 1\;{\rm eV}^2$ and $\sin^2 2 \theta_{\rm sol} \sim 1$.

The effective weighted sum  of Majorana neutrino masses contributing to the neutrinoless double $\beta $ decay $\langle m_e \rangle \equiv |\sum_i U^2_{\alpha i} m_i|$ is in our texture equal to ${\stackrel{0}{m}}$. Thus, the experimental upper limit for $\langle m_e \rangle $ gives $ {\stackrel{0}{m}} = \langle m_e \rangle < 0.4 (0.2)$ eV -- 1 (0.6) eV ({\it cf.} Baudis 99B in Ref. [5]). If {\it e.g.} $ {\stackrel{0}{m}} \sim 0.2$ eV corresponding to $\sin^2 2 \theta_{\rm atm} \sim 0.96$, then $\sin^2 2 \theta_{\rm LSND} \sim 8.8 \times 10^{-9}$, $\Delta m^2_{\rm LSND} \sim 4.0 \times 10^{-2}\;{\rm eV}^2$ and $\sin^2 2 \theta_{\rm sol} \sim 1$.

Very recently, a possible positive evidence of the neutrinoless double $\beta $ decay has been reported for the first time [10]. The proposed extimation is 0.05 eV $\leq \langle m_e \rangle \leq 0.84$ eV with the best fit $\langle m_e \rangle \sim 0.39 $ eV. Then, in our texture, for $ \stackrel{0}{m} = \langle m_e \rangle \sim (0.05 - 0.39 - 0.84)$ eV corresponding to $ \sin^2 2\theta_{\rm atm} \sim 0.63 - 0.99 - 0.998 $ one obtains $ \sin^2 2\theta_{\rm LSND} \sim 2.2\times 10^{-6} - 6.0\times 10^{-10} - 2.8\times 10^{-11}$, $ \Delta m^2_{\rm LSND} \sim (2.5\times 10^{-3} - 1.5\times 10^{-1} - 7.1\times 10^{-1}) \;{\rm eV}^2$ and $ \sin^2 2\theta_{\rm sol} \sim 0.998 - 1 - 1$. If this evidence is confirmed, we will be sure that $\nu_e$ is a Majorana neutrino and, moreover, we will gain the first experimental estimate of its mass scale. In the case such as in our texture, where neutrino masses $m_1\,,\,m_2\,,\,m_3$ are nearly degenerate, this scale shall be also the mass scale of Majorana neutrinos $\nu_\mu$  and $\nu_\tau$. The case of near degeneracy of $m_1 \,,\,m_2 \,,\,m_3$ is here supported by the considerably large best fit of the mass-squared scale, $\langle m_e \rangle^2 \sim (0.39)^2\;{\rm eV}^2 = 0.15 \;{\rm eV}^2$, distinctly larger than the mass-squared differences $ \Delta m^2_{21} \ll \Delta m^2_{32} \sim 3\times 10^{-3} \;{\rm eV}^2$.

\vspace{0.2cm}

\ni {\bf 4. Conclusions}

\vspace{0.2cm}

We presented in this note an effective texture for six Majorana neutrinos, three active and three (conventional) sterile, based on the $6\times 6$ mass matrix defined in Eqs. (3) and (5), and leading to the mixing matrix given in Eqs. (6) and (7), as well as to the mass spectrum (9) or (10). We conjectured that the Dirac $3\times 3$ component of such a neutrino  mass matrix (when the bimaximal mixing, specific for neutrinos, is transformed out unitarily) gets a {\it fermion universal form} (15) similar to the $3\times 3 $ mass matrix for charged leptons and $3\times 3$ mass matrices for up and down quarks, constructed previously with a considerable success [2,3].

This texture {\it predicts} reasonably oscillations of solar $\nu_e$'s in a form not inconsistent with  LMA solar solution, if the SuperKamiokande value of the mass-squared scale for atmospheric $\nu_\mu $'s is taken as an input. In both cases, neutrino oscillations are practically maximal. The proposed texture also {\it predicts} very small, perhaps unobservable, LSND effect with the oscillation amplitude of the order $O[10^{-11}\, ({\rm eV}/{\stackrel{0}{m}})^4]$ and the mass-squared scale of the order $O({\stackrel{0}{m}}^2) + O(10^{-5}\,{\rm eV}^2)$. If {\it e.g.} ${\stackrel{0}{m}} = O(10^{-1}\,{\rm eV}) - O(1{\rm eV})$ corresponding to $\sin^2 2 \theta_{\rm atm} = O(0.9) - O(1)$, then $\sin^2 2 \theta_{\rm LSND} = O(10^{-7}) - O(10^{-11})$, $\Delta m^2_{\rm LSND} = O(10^{-2}\,{\rm eV}^2) - O(1 \,{\rm eV}^2)$ and $\sin^2 2 \theta_{\rm sol} = O(1)$. 

The negative result of Chooz experiment imposes on the oscillation amplitude of LSND effect (in our texture) an upper bound of the order $O(10^{-3})$ which corresponds for ${\stackrel{0}{m}}$ to a lower limit of the order $O(10^{-2} \,{\rm eV})$ and for $\Delta m^2_{\rm LSND}$ to a lower limit of the order $O(10^{-4} \,{\rm eV}^2)$. Notice that the estimations following from the original LSND experiment [8] are {\it e.g.} $ \sin^2 2 \theta_{\rm LSND} = O(10^{-2})$ and $\Delta m^2_{\rm LSND} = O(1 \,{\rm eV}^2)$. The new miniBooNE experiment may confirm or revise the original LSND results.

As far as the neutrino mass spectrum is concerned, our model of neutrino texture is of 3 + 3 type, in contrast to the models of 3 + 1 or 2 + 2 types [11] discussed in the case when, beside three active neutrinos $\nu_e\,,\,\nu_\mu\,,\,\nu_\tau $, there is one {\it extra} sterile neutrino $\nu_s $. In those models, three Majorana {\it conventional} sterile neutrinos $\nu_{e_s}\,,\, \nu_{\mu_s} \,,\, \nu_{\tau_s}$ are decoupled through the familiar seesaw mechanism, as being practically identical with three very heavy neutrino mass states $\nu_4\,,\,\nu_5\,,\,\nu_6 $ (of the GUT mass scale). In our model, on the contrary, $\nu_{e_s}\,,\, \nu_{\mu_s} \,,\,\nu_{\tau_s}$ are practically identical with three mass states $\nu_4\,, \,\nu_5\,,\,\nu_6 $  that this time are constructed to be massless.

In this paper, the most crucial  may be the pertinent question, what is the physical (Higgs?) origin of the Dirac component $M^{(D)}$, Eq. (5), of the neutrino mass matrix $M$, where its bimaximal-mixing-free unitary transform ${\stackrel{0}{M}}\!\,^{(D)}$, Eq. (13), is conjectured to be of the fermion universal form (15) (with $\alpha/\mu $ negligible in the case of neutrinos). A somewhat different question arises also about the physical (explicit or effective?) origin of the lefthanded and righthanded components $ M^{(L)}$ and $ M^{(R)}$, Eqs. (5), of $M$.

The reader can find three Appendices added at the end of this paper. In Appendix A, an alternative, effective $6\times 6$ neutrino texture is sketched, where due to a specific degeneracy of the mass matrix there are no oscillations of the (conventional) sterile neutrinos, and, therefore, no LSND effect can arise. Appendix B contains a proposal of the explanation, why in nature there are three and only three generations of leptons and quarks, and also an argument for the particular form of the Dirac-type $3\times 3$ mass matrix used in this paper for neutrinos (and in Refs. [2] and [3] for charged leptons and quarks, respectively). Finally, Appendix C deals briefly with the problem of new boson hierarchy, appearing as an unavoidable by-product of explaining the observed fermion hierarchy in the way presented in Appendix B.

\vfill\eject

~~~~
\vspace{0.2cm}

\centerline{\bf Appendix A:}

\centerline{\bf An alternative $6\times 6$ texture without the LSND effect}

\vspace{0.35cm}

In this Appendix, we report on another effective texture for three active and three (conventional) sterile neutrinos, where there are {\it no oscillations} of the latter neutrinos due to a specific degeneracy of the mass matrix. Thus, they are decoupled from the former neutrinos, evidently in a different way than through the familiar seesaw mechanism.

In such a texture, the $3\times 3$ components of the neutrino $6\times 6$ mass matrix (3) get the form

$$
M^{(L)} = {\stackrel{0}{m}} \left( \begin{array}{ccc} 1 & 0 & 0 \\ 0 & 1 & 0 \\ 0 & 0 & 1 \end{array} \right) = - M^{(R)} \;,\;  M^{(D)} = {\stackrel{0}{m}}\left( \begin{array}{ccc} \frac{\tan 2\theta_{14}}{\sqrt{2}} & \frac{\tan 2\theta_{25}}{\sqrt{2}} & 0 \\ -\frac{\tan 2\theta_{14}}{2} &  \frac{\tan 2\theta_{25}}{2} & \frac{\tan 2\theta_{36}}{\sqrt{2}} \\ \frac{\tan 2\theta_{14}}{2} & -\frac{\tan 2\theta_{25}}{2} & \frac{\tan 2\theta_{36}}{\sqrt{2}} \end{array}\right) \eqno({\rm A.1})
$$

\vspace{0.15cm}

\ni with ${\stackrel{0}{m}} >0$ being a mass scale and $\tan 2\theta_{ij}\;\,(ij = 14,25,36)$ denoting three dimensionless parameters. Its unitary diagonalizing matrix is given as before in Eqs. (6) and (7), but now the relations

$$
c^2_{ij} - s^2_{ij} = \cos 2\theta_{ij} = \frac{1}{\sqrt{1 + \tan^2 2\theta_{ij}}} \eqno({\rm A.2})
$$

\ni work and the neutrino mass spectrum becomes

$$
m_{1,4} = \pm {\stackrel{0}{m}}\sqrt{1 + \tan^2 2\theta_{14}} \,,\, m_{2,5} = \pm {\stackrel{0}{m}}\sqrt{1 + \tan^2 2\theta_{25}} \,,\,m_{3,6} = \pm {\stackrel{0}{m}}\sqrt{1 + \tan^2 2\theta_{36}} \;, \eqno({\rm A.3})
$$

\ni satisfying the equalities

$$
\left(c^2_{14} - s^2_{14} \right) m_{1,4} = \left(c^2_{25} - s^2_{25}\right) m_{2,5} = \left(c^2_{36} - s^2_{36} \right) m_{3,6} = \pm {\stackrel{0}{m}} \;. \eqno({\rm A.4})
$$

\vspace{0.1cm}

\ni This can be seen by applying the formula $M_{\alpha \beta} = \sum_i U_{\alpha i} m_i U^*_{\beta i}\,$ with the use of mass spectrum described in Eqs. (A.3) or (A.4).

For the new texture, the neutrino oscillation formulae (20) lead to the relations 

\begin{eqnarray*}
P(\nu_e \rightarrow \nu_e)_{\rm sol}\;\;\;\, & = & 1 - \sin^2 (x _{21})_{\rm sol}\;,  \\
P(\nu_\mu \rightarrow \nu_\mu)_{\rm atm} \;\; & = & 1 - \frac{1}{4} \sin^2 (x _{21})_{\rm atm} - \frac{1}{2}\left[\sin^2 (x_{31})_{\rm atm} + \sin^2 (x_{32})_{\rm atm}\right]  \\
& \simeq & 1 - \sin^2 (x_{32})_{\rm atm} \;,  \\
P(\nu_\mu \rightarrow \nu_e)_{\rm LSND} & = & \frac{1}{2} \sin^2 (x _{21})_{\rm LSND} \simeq 0\;, \\
P(\bar{\nu}_e \rightarrow \bar{\nu}_e)_{\rm Chooz} & = & 1 - \sin^2 (x _{21})_{\rm Chooz} \simeq 1\;,  
\end{eqnarray*}

\vspace{-1.67cm}

\begin{flushright}
({\rm A}.5)
\end{flushright}

\vspace{0.1cm}

\ni where $0 \simeq (x_{21})_{\rm atm} \ll (x_{31})_{\rm atm} \simeq (x_{32})_{\rm atm}$, $(x_{21})_{\rm LSND} \simeq 0$ and $(x_{21})_{\rm Chooz} \simeq (x_{21})_{\rm atm} \simeq 0$. Note that the formulae (A.5) describe oscillations having the same form as those in the case of the simple bimaximal texture of three active neutrinos [12], but now with the specific mass spectrum (A.3). On the other hand, oscillations of three (conventional) sterile neutrinos vanish in the new texture,  $P(\nu_\alpha \rightarrow \nu_{\beta_s}) = 0$ and $P(\nu_{\alpha_s} \rightarrow \nu_{\beta_s}) = \delta_{\beta_s\,\alpha_s} \;\;(\alpha\,,\,\beta = e\,,\,\mu\,,\,\tau)$, in consequence of the degeneracy $ \Delta m_{41}^2 = \Delta m_{52}^2 = \Delta m_{63}^2 = 0 $ following from the equalities $m_1 = -m_4\;,\;m_2 = -m_5\;,\;m_3 = -m_6 $.

The oscillation formulae (A.5) imply {\it bimaximal mixing} for solar $\nu_e$'s and atmospheric $\nu_\mu$'s, {\it negative result} for Chooz reactor $\bar{\nu}_e$'s and {\it no} LSND {\it effect} for accelerator $\nu_\mu $'s ($\bar{\nu}_\mu $'s).

In the case of the conjecture (15) with ${\stackrel{0}{M}}\!\,^{(D)} = U^{(3) \dagger} M^{(D)}$, the new texture gives

$$
{\stackrel{0}{m}}\tan 2\theta_{14} = \frac{\mu}{29} \varepsilon \rightarrow 0 \,,\,{\stackrel{0}{m}} \tan 2\theta_{25} = \frac{\mu}{29} \frac{4\cdot 80}{9}  = 1.23 \mu \,,\,{\stackrel{0}{m}} \tan 2\theta_{36} = \frac{\mu}{29} \frac{24 \cdot 624}{25} = 20.7 \mu\,
\eqno({\rm A.6})
$$

\ni and then, from Eq. (A.3)

$$
m_{1,4} = \pm {\stackrel{0}{m}}\,,\, m_{2,5} = \pm \sqrt{{\stackrel{0}{m}}\,\!^2 + 1.50 \mu^2} \,,\, m_{3,6} = \pm \sqrt{ {\stackrel{0}{m}}\,\!^2 + 427 \mu^2}\;.\eqno({\rm A.7})
$$


\ni Hence,

$$
\Delta m^2_{21} = 1.50 \, \mu^2 \;,\; \Delta m^2_{32} = 425\, \mu^2 \;,\; \Delta m^2_{31} = 427\, \mu^2 \;. \eqno({\rm A.8})
$$

\ni Thus, using the SuperKamiokande result $ \Delta m^2_{32} \sim 3 \times 10^{-3}\;\,{\rm eV}^2 $ for atmospheric $\nu_\mu$'s described by the second formula (A.5), we obtain from Eq. (A.8)

$$ 
\mu^2 \sim 7.1 \times 10^{-6} \; {\rm eV}^2\;\, {\rm or}\,\;\mu \sim 2.7 \times 10^{-3} \; {\rm eV} \eqno({\rm A.9})
$$ 

\ni in place of the estimate (24). Then, from Eq. (A.8) we {\it predict}

$$ 
\Delta m^2_{21} \sim 1.1 \times 10^{-5}\;\,{\rm eV}^2 \eqno({\rm A.10})
$$ 

\ni for solar $\nu_e$'s presented in the first formula (A.5). So, the solar mass-squared scale  $\Delta m^2_{21}$ turns out to be the same as estimated before in Eq. (26), being not inconsistent with the LMA solar solution.

\vspace{0.7cm}

{\centerline{\bf Appendix B:}}

{\centerline{\bf Foundations for the fermion hierarchy}}

\vspace{0.4cm}

The form of Dirac mass matrix 

$$
{M}^{(f)} = \frac{1}{29} \left(\begin{array}{ccc} \mu^{(f)}\varepsilon^{(f)} & 2\alpha^{(f)}  & 0 
\\ 2\alpha^{(f)} & 4\mu^{(f)}(80 + \varepsilon^{(f)})/9 & 8\sqrt{3}\,\alpha^{(f)} 
\\ 0 & 8\sqrt{3} \,\alpha^{(f)} & 24\mu^{(f)} (624 + \varepsilon^{(f)})/25 \end{array}\right) \;, \eqno({\rm B}.1)
$$

\ni explored previously for charged leptons $ (f = e) $ [2] as well as for up and down quarks ($ f = u\,,\,d $) [3] with a considerable success, is applied in the present paper [Eq. (15)] to neutrinos ($f = \nu $), namely to the bimaximal-mixing-free unitary transform $\stackrel{0}{M}\!^{(D)}$ of Dirac component of their $6\times 6$ mass matrix $M$ ({\it cf.} also Ref. [4]). In this case, $\varepsilon^{(\nu)} \rightarrow 0$ and $\alpha^{(\nu)}/\mu^{(\nu)}$ is negligible. In consequence, $\Delta m^2_{\rm sol} = \Delta m^2_{21}$ is predicted just a little bit below the range suggested by the LMA solar solution, if the \UK result for $\Delta m^2_{\rm atm} = \Delta m^2_{32}$ is used. Notice that in the quark case ($f = u\,,\,d$) the parameter $\varepsilon^{(f)}$ must be replaced in the matrix element ${M}^{(f)}_{33}$ by $\varepsilon^{(f)} + C^{(f)}$, where $C^{(f)} > 0$ is large. 

In this Appendix, we argue, first of all, for there being three and only three generations of leptons and quarks, and then, for the particular form (B.1) of the Dirac-type mass matrix. This argumentation is based on two assumptions: 

\ni ({\it i}) the conjecture that all kinds of matter's fundamental particles existing in nature can be deduced from Dirac square-root procedure $ \sqrt{p^2} \rightarrow \Gamma \cdot p$, but constrained by an intrinsic Pauli principle, and 

\ni ({\it ii}) a simple specific ansatz for the shape of Dirac mass matrix, formulated on the ground of the first assumption.

\ni The conjecture ({\it i}) turns out to be sufficient to explain the puzzling existence of {\it three and only three} generations of leptons and quarks. Then, the ansatz ({\it ii}) reproduces the specific form (B.1) of the Dirac mass matrix. At the end of this Appendix, we speculate on the physical origin of the ansatz ({\it ii}).

It is not difficult to see that, in the interaction-free case, Dirac's square-root procedure implies generically the sequence $N = 1,2,3,\ldots $ of generalized Dirac equations [13,2]:

$$
\left( \Gamma^{(N)}\cdot p -  M^{(N)}\right) \psi^{(N)}(x) = 0\,, 
\eqno({\rm B}.2)
$$

\ni where for any $N$ the Dirac algebra 

\vspace{-0.1cm}

$$
\left\{ \Gamma^{(N)}_\mu\,,\,\Gamma^{(N)}_\nu \right\} = 2 g_{\mu \nu}
\eqno({\rm B}.3)
$$

\ni  holds, constructed by means of a Clifford algebra:

$$
\Gamma^{(N)}_\mu \equiv \frac{1}{\sqrt{N}} \sum^N_{i=1}  \gamma^{(N)}_{i \mu}\;\; , \;\;\left\{ \gamma^{(N)}_{i \mu}\,,\,\gamma^{(N)}_{j \nu} \right\} = 2 \delta_{i j} g_{\mu \nu}
\eqno({\rm B}.4)
$$

\ni with $i\,,\,j = 1,2,\ldots ,N$ and $\mu\,,\,\nu = 0,1,2,3$. The mass $ M^{(N)}$ is independent of $\Gamma^{(N)}_\mu$. In general, the mass $ M^{(N)}$ should be replaced by a mass matrix of elements $ M^{(N,N')}$ which would couple $\psi^{(N)}(x)$ with all appropriate $\psi^{(N')}(x)$, and it might be natural to assume for $N \neq N'$ that $\gamma^{(N)}_{i \mu}$ and $\gamma^{(N')}_{j \nu}$ commute, and so do $\Gamma^{(N)}_\mu $ and $\Gamma^{(N')}_\nu $.

For $ N = 1$, Eq. (B.2) is evidently the usual Dirac equation and for $ N = 2$ it is known as the Dirac form [14] of  K\"{a}hler equation [15], while for $ N \geq 3$ Eqs. (B.2) give us {\it new} Dirac-type equations [13,2]. They describe some spin-halfinteger or spin-integer particles for $N$ odd or $N$ even, respectively. 

The Dirac-type matrices $\Gamma^{(N)}_\mu $ for any $N$ can be embedded into the new Clifford algebra

\vspace{-0.1cm}

$$
\left\{ \Gamma^{(N)}_{i \mu}\, ,  \,\Gamma^{(N)}_{j \nu} \right\} = 2\delta_{i j} g_{\mu \nu}\;,
\eqno({\rm B}.5)
$$

\ni isomorphic with the Clifford algebra of $\gamma^{(N)}_{i \mu}$, if $\Gamma^{(N)}_{i \mu}$ are defined by the properly normalized Jacobi linear combinations of $\gamma^{(N)}_{i \mu}$:

\begin{eqnarray*}
\Gamma^{(N)}_{1 \mu} & \equiv & \Gamma^{(N)}_\mu \equiv \frac{1}{\sqrt{N}} \sum^N_{i=1}   \gamma^{(N)}_{i \mu}\, , \\ \Gamma^{(N)}_{i \mu} & \equiv & \frac{1} {\sqrt{i(i - 1)}} \left[ \gamma^{(N)}_{1 \mu} + \ldots + \gamma^{(N)}_{i\!-\!1\, \mu} - (i - 1) \gamma^{(N)}_{i \mu} \right] 
\end{eqnarray*}

\vspace{-1.6cm}

\begin{flushright}
({\rm B}.6)
\end{flushright}


\ni for $ i = 1$ and $ i = 2,\ldots, N$, respectively. So, $\Gamma^{(N)}_{1 \mu}$ and $ \Gamma^{(N)}_{2 \mu} \, , \, \ldots,\Gamma^{(N)}_{N \mu}$, respectively, present  the  "centre-of-mass" \,and "relative" \,Dirac-type matrices. Note that the Dirac-type equation (B.2) for any $N$ does not involve the "relative" \,Dirac-type matrices $\Gamma^{(N)}_{2 \mu} \, , \, \ldots, \Gamma^{(N)}_{N \mu}$, including solely the "centre-of-mass" \,Dirac-type matrix $\Gamma^{(N)}_{1 \mu} \equiv \Gamma^{(N)}_\mu $. Since $\Gamma^{(N)}_{i \mu} = \sum^N_{j=1} O_{i j} \gamma^{(N)}_{j \mu}$, where $ O = \left( O_{i j} \right)$ is an orthogonal $N\times N$ matrix ($O^T = O^{-1}$), we obtain for the total spin tensor the equality

$$
\sum^N_{i=1}  \sigma^{(N)}_{i \mu \nu} = \sum^N_{i=1}  \Sigma^{(N)}_{i \mu \nu}  \,,
\eqno({\rm B}.7)
$$

\ni where

$$ 
\sigma^{(N)}_{j \mu \nu} \equiv \frac{i}{2} \left[ \gamma^{(N)}_{j \mu} \, , \, \gamma^{(N)}_{j \nu} \right] \;\; ,\;\; \Sigma^{(N)}_{j \mu \nu} \equiv \frac{i}{2} \left[ \Gamma^{(N)}_{j \mu} \, , \, \Gamma^{(N)}_{j \nu} \right] \,. \eqno({\rm B}.8)
$$

\ni The total spin tensor (B.7) is the generator of Lorentz transformations for $\psi^{(N)}(x)$.

In place of the chiral representations for individual $\gamma^{(N)}_ j = \left(\gamma^{(N)}_{j \mu} \right)$, where   

\vspace{-0.1cm}

$$
\gamma^{(N)}_{j 5} \equiv i \gamma^{(N)}_{j 0} \gamma^{(N)}_{j 1} \gamma^{(N)}_{j 2} \gamma^{(N)}_{j 3} \;\; , \;\; \sigma^{(N)}_{j 3} \equiv \sigma^{(N)}_{j 12}
\eqno({\rm B}.9)
$$

\ni are diagonal, it is convenient to use for any $N$ the chiral representations of Jacobi $\Gamma^{(N)}_j = \left(\Gamma^{(N)}_{j \mu} \right)$, where now

$$
\Gamma^{(N)}_{j 5} \equiv i\Gamma^{(N)}_{j 0} \Gamma^{(N)}_{j 1} \Gamma^{(N)}_{j 2} \Gamma^{(N)}_{j 3} \;\; ,\;\; \Sigma^{(N)}_{j 3} \equiv  \Sigma^{(N)}_{j 12} \eqno({\rm B}.10)
$$

\vspace{0.15cm}
\ni are diagonal (all matrices (B.9) and similarly (B.10) commute simultaneously, both with equal and different $j$). 

When using the Jacobi chiral representations, the "centre-of-mass" \,Dirac-type matrices  $\Gamma^{(N)}_{1 \mu} \equiv \Gamma^{(N)}_\mu$ and $ \Gamma^{(N)}_{1 5} \equiv \Gamma^{(N)}_5 \equiv i\Gamma^{(N)}_0 \Gamma^{(N)}_1 \Gamma^{(N)}_2 \Gamma^{(N)}_3$ can be taken in the reduced forms

$$
\Gamma^{(N)}_\mu =  \gamma_\mu \otimes \underbrace{ {\bf 1}\otimes \cdots \otimes {\bf 1}}_{ N-1 \;{\rm times}} \;\; , \;\; \Gamma^{(N)}_5 = \gamma_5  \otimes \underbrace{ {\bf 1}\otimes \cdots \otimes {\bf 1}}_{ N-1 \;{\rm times}} \; , 
\eqno({\rm B}.11)
$$

\ni where $\gamma_\mu$, $ \gamma_5 \equiv i \gamma_0 \gamma_1 \gamma_2 \gamma_3 $ and {\bf 1} are the usual $4\times 4$ Dirac matrices. 

Then, the Dirac-type equation (B.2) for any $N$ can be rewritten in the reduced form

$$
\left( \gamma \cdot p  - M^{(N)}\right)_{\alpha_1\beta_1} \psi^{(N)}_{\beta_1 \alpha_2 \ldots \alpha_N}(x) = 0\;,
\eqno({\rm B}.12)
$$

\ni where $\alpha_1$ and $\alpha_2 \,,\, \ldots\,,\, \alpha_N$ are the "centre-of-mass" \,and "relative" \,Dirac bispinor indices, respectively ($\alpha_i =1,2,3,4$ for any $i = 1,2,\ldots,N$). Note that in the Dirac-type equation (B.12) for any $N>1$ there appear the "relative" \,Dirac indices $\alpha_2 \,,\, \ldots\,,\, \alpha_N$ which are free from any coupling, but still are subjects of Lorentz transformations. 

The Standard Model gauge interactions can be introduced to the Dirac-type equations (B.12) by means of the minimal substitution $p \rightarrow p - g A(x)$, where $p$ plays the role of the "centre-of-mass" \,four-momentum, and so, $x$ --- the "centre-of-mass" \,four-position. Then,

$$
\left\{ \gamma \cdot \left[p - g A(x)\right] - M^{(N)}\right\}_{\alpha_1\beta_1} \psi^{(N)}_{\beta_1 \alpha_2 \ldots \alpha_N}(x) = 0\;,
\eqno({\rm B}.13)
$$

\ni where $g \gamma \cdot A(x)$ symbolizes the Standard Model gauge coupling that involves within $A(x)$ the familiar weak-isospin and color matrices as well as the usual Dirac chiral matrix $\gamma_5 $. The last arises from the "centre-of-mass" \, Dirac-type chiral matrix $\Gamma^{(N)}_5 $, when a generic $g \Gamma^{(N)} \cdot A(x)$ is reduced to $g \gamma \cdot A(x)$  in Eqs. (B.13) [see Eq. (B.11)].

In Eqs. (B.13) the Standard Model gauge fields interact only with the "centre-of-mass" \,index $\alpha_1$ that, therefore, is distinguished from the physically unobserved "relative" \,indices $\alpha_2 \,,\, \ldots\,,\, \alpha_N$. This was the reason, why some time ago we conjectured that the "relative" \,Dirac bispinor  indices $\alpha_2 \,,\, \ldots\,,\, \alpha_N$ are all undistinguishable physical objects obeying Fermi statistics along with the Pauli principle requiring the full antisymmetry of wave function $\psi^{(N)}_{\alpha_1 \alpha_2 \ldots \alpha_N}(x)$ with respect to $\alpha_2 \,,\, \ldots\,,\, \alpha_N$ [13,2]. Hence, due to this "intrinsic Pauli principle", only five values of $N$ satisfying the condition $N-1\leq 4$ are allowed, namely $N = 1,3,5$ for $N$ odd and $N = 2,4$ for $N$ even. Then, from the postulate of relativity and the probabilistic interpretation of $\psi^{(N)}(x) = \left(\psi^{(N)}_{\alpha_1 \alpha_2 \ldots \alpha_N}(x)\right) $ we were able to infer that these $N$ odd and $N$ even correspond to states with total spin 1/2 and total spin 0, respectively [13,2].

Thus, the Dirac-type equation (B.13), jointly with the intrinsic Pauli principle, if considered on a fundamental level, justifies the existence in nature of {\it three and only three} generations of spin-1/2 fundamental fermions coupled to the \SM gauge bosons  (they are identified with leptons and quarks). In addition, there should exist {\it two and only two} generations of spin-0 fundamental bosons also coupled to the \SM gauge bosons (they are not identified yet).

The wave functions or fields of spin-1/2 fundamental fermions (leptons and quarks) of three generations $N = 1,3,5$ can be presented in terms of $\psi^{(N)}_{\alpha_1 \alpha_2 \ldots \alpha_N}(x)$ as follows:

\vspace{-0.2cm} 

\begin{eqnarray*} 
\psi^{(f_1)}_{\alpha_1}(x) & = & \psi^{(1)}_{\alpha_1}(x) \;, \nonumber \\
\psi^{(f_3)}_{\alpha_1}(x) & = & \frac{1}{4}\left(C^{-1} \gamma_5 \right)_ {\alpha_2 \alpha_3} \psi^{(3)}_{\alpha_1 \alpha_2 \alpha_3}(x) = \psi^{(3)}_{\alpha_1 1 2}(x) = \psi^{(3)}_{\alpha_1 3 4}(x) \;,\nonumber \\
\psi^{(f_5)}_{\alpha_1}(x) & = & \frac{1}{24}\varepsilon_{\alpha_2 \alpha_3 \alpha_4 \alpha_5} \psi^{(5)}_{\alpha_1 \alpha_2 \alpha_3 \alpha_4 \alpha_5}(x) = \psi^{(5)}_{\alpha_1 1 2 3 4}(x) \;,
\end{eqnarray*}  

\vspace{-1.65cm}

\begin{flushright}
(B.14)
\end{flushright}

\vspace{0.1cm} 

\ni where $ \psi^{(N)}_{\alpha_1 \alpha_2 \ldots \alpha_N}(x) $ carries also the \SM (composite) label, suppressed in our notation, and $C$ denotes the usual $4\times 4$ charge-conjugation matrix. Here, writing explicitly, $f_1 = \nu_e\,,\,e^-\,,\, u\,,\,d \;,\; f_3 = \nu_\mu\,,\, \mu^-\,,\, c\,,\,s $ and $f_5 = \nu_\tau\,,\, \tau^-\,,\,t \,,\,b $, thus each $f_N$ corresponds to the same suppressed \SM (composite) label. We can see that, due to the full antisymmetry in $\alpha_i $ indices for $i \geq 2$, the wave functions or fields $N =1,3$ and 5 appear (up to the sign) with the multiplicities 1, 4 and 24,  respectively. Thus, for them, there is defined the weighting matrix

$$ 
\rho^{1/2} = \frac{1}{\sqrt{29}}  \left( \begin{array}{ccc}  1  & 0 & 0 \\ 0 & \sqrt4 & 0  \\ 0 & 0 & \sqrt{24} \end{array} \right) \;, \eqno({\rm B}.15)
$$

\ni where Tr $\rho = 1$.

Concluding the first part of this Appendix, we would like to point out that our algebraic construction 
of {\it three and only three} generations of leptons and quarks may be interpreted {\it either} as ingenuously algebraic (much like the famous Dirac's algebraic discovery of spin 1/2), {\it or} as a summit of an iceberg of really composite states of $N$ spatial partons with spin 1/2 whose Dirac bispinor indices manifest themselves as our Dirac bispinor  indices $\alpha_1 \,,\, \alpha_2 \,,\, \ldots\,,\, \alpha_N$ ($N = 1,3,5$) which thus may be called "algebraic partons", as being algebraic building blocks for leptons and quarks. Among all $N$ "algebraic partons" \,in any generation $N$ of leptons and quarks, there are one "centre-of-mass algebraic parton" \,$ (\alpha_1)$ and $ N-1$ "relative algebraic partons" \,$(\alpha_2 \,,\,\ldots\,,\,\alpha_N)$, the latter undistinguishable from each other and so, obeying our intrinsic Pauli principle.  

In the second part of this Appendix we introduce a simple specific ansatz for the shape of Dirac mass matrix by putting [13,2]

$$
M^{(f)} = \rho^{1/2} h^{(f)} \rho^{1/2} \, , \eqno({\rm B}.16)
$$

\ni where $\rho^{1/2}$ is given in Eq. (B.15) and

$$ 
h^{(f)} =  \mu^{(f)}\left[ N^2 - (1 - \varepsilon^{(f)}) N^{-2}\right] + \alpha^{(f)}(a  + a^\dagger)
\eqno({\rm B}.17)
$$

\ni with $\mu^{(f)} > 0$ and $\varepsilon^{(f)} > 0$ being parameters, while $f = \nu\,,\, e\,,\,u\,,\,d$ refers to neutrinos, charged leptons, up quarks and down quarks, respectively. Here, the matrix

$$ 
N  = \left( \begin{array}{ccc} 1 & 0 & 0 \\ 0 & 3 & 0 \\ 0 & 0 & 5 \end{array}\right) = 1 + 2 n  
\eqno({\rm B}.18)
$$

\ni describes the number of all $\alpha_i $ indices with $i = 1,2,\ldots,N$ (all "algebraic partons"), appearing in any of three fermion generations $N = 1,3,5$, while

$$ 
a = \left( \begin{array}{ccc} 0 & 1 & 0 \\ 0 & 0 & \sqrt2 \\ 0 & 0 & 0 \end{array}\right)  \; ,\;a^\dagger = \left( \begin{array}{ccc} 0 & 0 & 0 \\ 1 & 0 & 0 \\ 0 & \sqrt2 & 0 \end{array}\right)  
\eqno({\rm B}.19)
$$

\ni play the role of "truncated" \,annihilation and creation matrices for pairs of "relative" \,indices $\alpha_i \alpha_j $ with $ i,j = 2,\ldots,N$ (pairs of "relative algebraic partons"):

$$ 
[a\, , \,n] = a \;,\; [a^\dagger\, , \,n] = -a^\dagger\;,\;n = a^\dagger a = \left( \begin{array}{ccc} 0 & 0 & 0 \\ 0 & 1 & 0 \\ 0 & 0 & 2 \end{array}\right) \, ,  
\eqno({\rm B}.20)
$$

\ni where the "truncation" \,condition $ a^3 = 0 = a^{\dagger \,3}$ is satisfied. 

It is not difficult to show that the formulae (B.16) and (B.17) lead explicitly to the particular form (B.1) of Dirac-type mass matrix.

Finally, a few words about a possible physical origin of the ansatz (B.17). In the kernel  (B.17) of the Dirac mass matrix (B.16), the first term $\mu^{(f)} N^2$ may be intuitively interpreted as coming from an interaction of all $N$ "algebraic partons" \,treated on equal footing, while the second term $-\mu^{(f)} (\!1 - \!\varepsilon^{(f)}) N^{-2}$ may be considered as being a subtraction term caused by the fact that there is one "centre-of-mass algebraic parton" \,distinguished (due to its external coupling to the \SM gauge fields) among all $N$ "algebraic partons" \,of which $N\!-\!1$ are "relative algebraic partons", undistinguishable from each other.  This distinguished "algebraic parton" \,appears, therefore, with the probability $ [N!/(N\!-\!1)!]^{-1} = N^{-1}$ that, when squared, leads to the additional term $\mu^{(f)}(1 - \varepsilon^{(f)}) N^{-2}$ [with a coefficient $ \mu^{(f)}(1 - \varepsilon^{(f)})$] which should be subtracted in the kernel (B.17) from the former term in order to obtain the mass matrix element $M^{(f)}_{11} = \mu^{(f)} \varepsilon^{(f)}/29$ tending to zero if $\varepsilon^{(f)} \rightarrow 0$. Eventually, the third term $\alpha^{(f)} (a + a^\dagger)$ in the kernel (B.17)  annihilates and creates pairs of "relative algebraic partons" \,and so, is responsible in a natural way for mixing of three fermion generations in the Dirac mass matrix $M^{(f)}$.

\vspace{0.7cm}

{\centerline{\bf Appendix C:}}

{\centerline{\bf Problem of new boson hierarchy}}

\vspace{0.4cm}

The way of explanation of the observed fermion hierarchy (especially, of the existence of three and only three generations of leptons and quarks), as is described in Appendix B, suggests also the existence of a new boson hierarchy, consisting of {\it two and only two} generations of spin-0 fundamental bosons. These boson generations correspond to the numbers $N = 2,4$ of the Dirac bispinor indices $\alpha_1 \,,\, \alpha_2 \,,\, \ldots \,,\, \alpha_N$, among which there are one "centre-of-mass" \,\,index $\alpha_1$ and $N-1 = 1,3$ "relative" \,indices $\alpha_2$ or $\alpha_2 \,,\, \alpha_3 \,,\, \alpha_4$, respectively. Only the "centre-of-mass" \,index $\alpha_1$ is coupled to the \SM gauge bosons.

The wave functions or fields of spin-0 fundamental bosons of two generations $N = 2,4$ can be written down in terms of $\psi^{(N)}_{\alpha_1\alpha_2\ldots \alpha_N} (x)$ as follows:

\vspace{-0.4cm}
 
\begin{eqnarray*}
\psi^{(b_2)}(x) & = & \frac{1}{2\sqrt{2}}(C^{-1}\gamma_5)_{\alpha_1\alpha_2} \psi^{(2)}_{\alpha_1 \alpha_2}(x) \\ & =  & \frac{1}{\sqrt{2}} \left[\psi^{(2)}_{12}(x) - \psi^{(2)}_{21}(x) \right] = \frac{1}{\sqrt{2}} \left[\psi^{(2)}_{34}(x) - \psi^{(2)}_{43}(x) \right] \;, \\ 
\psi^{(b_4)}(x) & = & \frac{1}{6\sqrt{4}} \varepsilon_{\alpha_1\alpha_2\alpha_3
\alpha_4}\psi^{(4)}_{\alpha_1\alpha_2\alpha_3\alpha_4} (x) \\ & = & \frac{1}{\sqrt{4}} \left[\psi^{(4)}_{1234}(x) - \psi^{(4)}_{2134}(x) + \psi^{(4)}_{3412}(x) - \psi^{(4)}_{4312}(x) \right] \;,
\end{eqnarray*} 

\vspace{-1.57cm}

\begin{flushright}
(C.1)
\end{flushright}

\vspace{0.1cm} 

\ni where the wave function or field $\psi^{(N)}_{\alpha_1\alpha_2\ldots \alpha_N} (x)$ carries the suppressed \SM (composite) label. In consequence, there are four sorts of fundamental scalars carrying the same \SM signature as four sorts of fundamental fermions, namely as neutrinos ($f = \nu$), charged leptons ($f = e$), up quarks ($f  = u$) and down quarks ($f = d$). These fermions, however, are realized in three generations $N = 1,3,5$, while the fundamental scalars are predicted in two generations $N = 2,4$. So, one cannot hope here for a construction of the full supersymmetry (at most, there might appear a partial supersymmetry: two to two).

Two lepton-like scalar doublets (corresponding, as far as the \SM signature is concerned, to three lepton doublets) might play the role of two generations of Higgs doublets [16]. On the other hand, two quark-like scalar doublets (corresponding to three quark doublets) should lead to a lot of new (colorless) hadrons, composed dynamically of these colored scalars from two generations and (also colored) quarks from three generations [17]. Most of them should be highly unstable, but perhaps not all, allowing then for some new observations.

\vfill\eject

\baselineskip 0.72cm

\vspace{0.1cm}

{\centerline{\bf References}}

\vspace{0.15cm}

{\everypar={\hangindent=0.7truecm}
\parindent=0pt\frenchspacing

[1]~~For a theoretical summary {\it cf.} J.~Ellis, {\it Nucl. Phys. Proc. Suppl.} {\bf 91}, 503 (2001); and references therein.

[2]~~W.~Kr\'{o}likowski, in {\it Spinors, Twistors, Clifford Algebras and Quantum Deformations (Proc. 2nd Max Born Symposium 1992)}, eds. Z.~Oziewicz {\it et al.}, Kluwer Acad. Press, 1993; {\it Acta Phys. Pol.} {\bf B 27}, 2121 (1996); and references therein.

[3]~~W. Kr\'{o}likowski, {\it Acta Phys. Pol.} {\bf B 30}, 2631 (1999); hep--ph/0108157; and references therein.

[4]~~W. Kr\'{o}likowski, hep--ph/0109212.

[5]~~The Particle Data Group, {\it Eur. Phys. J.} {\bf C 15}, 1 (2000).

[6]~~T. Kajita and Y. Totsuka, {\it Rev. Mod. Phys.} {\bf 73}, 85 (2001); T. Toshito, hep--ex/0105023.

[7]~For a recent analysis {\it cf.} M.V.~Garzelli and C.~Giunti, hep--ph/0108191; hep--ph/0111254; {\it cf.} also V. Barger, D. Marfatia and K. Whisnant, hep--ph/0106207.

[8]~~G. Mills, {\it Nucl. Phys. Proc. Suppl.} {\bf 91}, 198 (2001);  R.L.~Imlay, Talk at {\it ICHEP 2000} at Osaka; and references therein.

[9]~~M. Appolonio {\it et al.}, {\it Phys. Lett.} {\bf B 420}, 397 (1998); {\bf B 466}, 415 (1999).

[10]~$\!$H.V. Klapdor-Kleingrothaus {\it et al.}, {\it Mod. Phys. Lett.} {\bf A 37}, 2409 (2002) (hep--ph/0201231).

[11]~{\it Cf. e.g.} V. Barger, B. Kayser, J. Learned, T. Weiler and K. Whisnant, {\it  Phys. Lett.} {\bf B 489}, 345 (2000); M.C. Gonzalez--Garcia, M. Maltoni and C. Pe\~{n}a--Garay, hep--ph/0108073; and references therein; {\it cf}. also W. Kr\'{o}likowski, hep--ph/0106350; O.Yasuda, hep--ph/0109067.

[12]~{\it Cf. e.g.} W. Kr\'{o}likowski, hep--ph/0007255.

[13]~W. Kr\'{o}likowski, {\it Acta Phys. Pol.} {\bf B 21}, 871 (1990); {\it Phys. Rev.} {\bf D 45}, 3222 (1992); {\it Acta Phys. Pol.} {\bf B 24}, 1149 (1993); {\it cf.} also Appendix in hep-ph/0108157 (the second Ref. [3]). 

[14]~T. Banks, Y. Dothan and D.~Horn, {\it Phys. Lett.} {\bf B 117}, 413 (1982).

[15]~E. K\"{a}hler, {\it Rendiconti di Matematica} {\bf 21}, 425 (1962); {\it cf.} also D.~Ivanenko and L.~Landau, {\it Z. Phys.} {\bf 48}, 341 (1928).

[16]~W. Kr\'{o}likowski, {\it Phys. Rev.} {\bf D 46}, 5188 (1992). 

[17]~W. Kr\'{o}likowski, {\it Acta Phys. Pol.} {\bf B 24}, 1149 (1993); {\bf B 26}, 1217 (1995); {\it Nuovo Cim.} {\bf 107 A}, 69 (1994).

\vfill\eject

\end{document}